\DeclareFontFamily{T1}{aer}{}
\DeclareFontShape{T1}{aer}{m}{n}{
  <-6> cmr5 <6-7> cmr6 <7-8> cmr7 <8-9> cmr8 <9-10>
       cmr9 <10-12> cmr10 <12-17> cmr12 <17-> cmr17 }{}
\DeclareFontShape{T1}{aer}{bx}{n}{
  <-6> cmbx5 <6-7> cmbx6 <7-8> cmbx7 <8-9> cmbx8 <9-10>
       cmbx9 <10-12> cmbx10 <12-> cmbx12 }{}
\DeclareFontShape{T1}{aer}{m}{it}{
  <-8> cmti7 <8-9> cmti8 <9-10> cmti9 <10-12> cmti10 <12-> cmti12 }{}
\DeclareFontShape{T1}{aer}{m}{sl}{
  <-9> cmsl8 <9-10> cmsl9 <10-12> cmsl10 <12-> cmsl12 }{}
\DeclareFontShape{T1}{aer}{b}{n}{ <-> cmb10 }{}
\DeclareFontShape{T1}{aer}{bx}{it}{ <-> cmbxti10 }{}
\DeclareFontShape{T1}{aer}{bx}{sl}{ <-> cmbxsl10 }{}
\DeclareFontShape{T1}{aer}{m}{sc}{ <-> cmcsc10 }{}
\DeclareFontShape{T1}{aer}{l}{n}{<->ssub * aer/m/n}{}
\DeclareFontShape{T1}{aer}{l}{it}{<->ssub * aer/m/it}{}
\DeclareFontShape{T1}{aer}{l}{sl}{<->ssub * aer/m/sl}{}
\DeclareFontShape{T1}{aer}{b}{it}{<->ssub * aer/bx/it}{}
\DeclareFontShape{T1}{aer}{b}{sl}{<->ssub * aer/bx/sl}{}
\DeclareFontShape{T1}{aer}{l}{sc}{<->ssub * aer/m/sc}{}

\DeclareFontFamily{T1}{aett}{}
\DeclareFontShape{T1}{aett}{m}{n}{
  <-6> cmr5 <6-7> cmr6 <7-8> cmr7 <8-9> cmr8 <9-10>
       cmr9 <10-12> cmr10 <12-17> cmr12 <17-> cmr17 }{}
\DeclareFontShape{T1}{aett}{m}{it}{
  <-6> cmr5 <6-7> cmr6 <7-8> cmr7 <8-9> cmr8 <9-10> cmr9
  <10-12> cmr10 <12-17> cmr12 <17-> cmr17 }{}
\documentclass{acm_proc_article-sp}

\usepackage{stmaryrd} % llbracket, rrbracket
\usepackage{url}      % url
\usepackage{theorem}  % theorembodyfont

\allowdisplaybreaks

\theorembodyfont{\normalfont}
\newtheorem{theorem}{Theorem}

\newtheorem{lemma}[theorem]{Lemma}
\newtheorem{corollary}[theorem]{Corollary}
\newdef{definition}{Definition}

\newcommand{\tup}[1]{\langle {#1} \rangle}
\newcommand{\sem}[1]{\llbracket {#1} \rrbracket}
\newcommand{\op}[1]{\operatorname{#1}}
\newcommand{\bigsem}[1]{\big\llbracket {#1} \big\rrbracket}
\newcommand{\abs}[1]{\lvert {#1} \rvert}
\newcommand{\nat}{\mathbb{N}}
\newcommand{\rhs}{{\rm rhs}}
\newcommand{\comp}{{\,;\,}}

\newcommand{\IO}{{\rm IO}}
\newcommand{\OI}{{\rm OI}}
\newcommand{\MTT}{{\rm MTT}}
\newcommand{\MTTOI}{\MTT_\OI}
\newcommand{\MTTIO}{\MTT_\IO}
\newcommand{\LMTT}{{\rm LMTT}}
\newcommand{\LMTTOI}{\LMTT_\OI}

\newcommand{\DMTT}{{\rm D}\MTT}

\newcommand{\DtMTT}{{\rm D}{\rm t}\MTT}
\newcommand{\LDtMTT}{{\rm LD}{\rm t}\MTT}
\newcommand{\YIELD}{{\rm YIELD}}

\newcommand{\rankfn}{\mbox{\it rank\/}}
\newcommand{\labelfn}{\mbox{\it label\/}}
\newcommand{\posfn}{\mbox{\it pos\/}}
\newcommand{\childfn}{\mbox{\it child\/}}

\newcommand{\syma}{{\tt a}}
\newcommand{\symb}{{\tt b}}
\newcommand{\symc}{{\tt c}}
\newcommand{\symd}{{\tt d}}
\newcommand{\syme}{{\tt e}}
\newcommand{\symv}{{\tt v}}

\begin{document}

\toappear{
  Copyright is held by the author/owner(s).\par
  International Workshop on Programming Language Techniques for XML (PLAN-X 2009),\par
  January 24, 2009, Savannah, Georgia.
}

\title{
  The Complexity of Translation Membership \\ for Macro Tree Transducers
}

\numberofauthors{2}

\author{
 \alignauthor
   Kazuhiro Inaba
   \\ \affaddr{The University of Tokyo}
   \\ \email{kinaba@is.s.u-tokyo.ac.jp}
 \alignauthor
   Sebastian Maneth
   \\ \affaddr{NICTA and University of New South Wales}
   \\ \email{sebastian.maneth@nicta.com.au}
}

\maketitle

%%%%%%%%%%%%%%%%%%%%%%%%%%%%%%%%%%%%%%%%%%%%%%%%%%%%%%%%%%%%%%%%%%%%%%%%%%%%%%%%

\begin{abstract}
Macro tree transducers (mtts) are
a useful formal model for XML query and transformation languages.
In this paper one of the fundamental decision problems on translations,
namely the ``translation membership problem'' is studied for mtts.
For a fixed translation, the translation membership problem asks whether
a given input/output pair is element of the translation.
For call-by-name mtts this problem is shown to be NP-complete. 
The main result is that translation membership for
call-by-value mtts is in polynomial time.
For several extensions, such as addition of regular
look-ahead or the generalization to multi-return mtts,
it is shown that translation membership still remains in PTIME.
\end{abstract}

%%%%%%%%%%%%%%%%%%%%%%%%%%%%%%%%%%%%%%%%%%%%%%%%%%%%%%%%%%%%%%%%%%%%%%%%%%%%%%%%

\section{Introduction}
\label{sec:introduction}

Macro tree transducers (mtts)~\cite{EV85} are a popular formal model
for XML query and transformation languages
(cf., e.g.,~\cite{EM03,MBPS05,MPS07}.
They are powerful enough to represent a wide range of practical
transformations, 
and they subsume various well-known models of tree translations such as
attribute grammars,
MSO-definable tree translations~\cite{Courcelle94},
or pebble tree transducers~\cite{MSV03}.
Yet, mtts have many decidable properties such as exact typechecking or
emptiness and finiteness and membership of their domains and ranges.
These make mtts a useful device for static verification of XML
translation programs.

In the algorithms that decide such properties, we sometimes encounter 
as a sub-problem the 
``translation membership problem''~\cite{IM08}. 
For a fixed translation, the translation membership problem asks whether
a given input/output pair is element of the translation.
Although the problem itself seems simple, it is far beyond trivial to solve
the problem efficiently, in particular if we consider nondeterministic mtts.
Nondeterminism is useful when using the mtt to 
approximate the behavior of a ``real'' (Turing-complete) 
programming language (viz. a complicated 
if-then-else expression; it is translated into an mtt that
nondeterministically chooses one of the conditional branches).
Depending on the order of evaluation, there are two different models of
nondeterministic mtts, namely, call-by-value (also called inside-out or
IO for short) and call-by-name (outside-in or OI).
Note that in the limit, 
to one given input tree of size $n$
an mtt can associate at most
$2^{2^{2^n}}$-many different output trees, if the
mtt operates in OI mode.
In contrast, the limit for mtts in IO mode is
at most $2^{2^n}$ different output trees for
a given input tree of size $n$.
Consider the following four rules of an mtt.
\[
  \begin{array}{lcl}
    \op{start} ({\tt a}(x_1))   & \to & \op{double}(x_1, \op{double}(x_1, {\tt e}))
  \\
    \op{double}({\tt a}(x_1),y_1) & \to & \op{double}(x_1, \op{double}(x_1, y_1))
  \\
    \op{double}({\tt e},y_1)    & \to & {\tt f}(y_1,y_1) \mid {\tt g}(y_1,y_1).
  \end{array}
\]
For an input tree of the form
$s_n={\tt a}({\tt a}(\cdots {\tt a}({\tt e})\cdots))$ with $n$ ${\tt a}$-nodes, this mtt
generates a full binary tree of height $2^n$ (and thus of
size $2^{2^n}$).
If the mtt operates in OI derivation mode, then each node
of the binary output tree is nondeterministically 
labeled either ${\tt f}$ or ${\tt g}$;
thus, there are $2^{2^{2^n}}$-many output trees associated
to the input tree $s_n$.
If, however, the mtt with the same rules operates in 
IO derivation mode, then for input $s_n$ it generates only 
$2^{2^n}$ many different output trees (the
nodes on one level of an output tree all have the same label).
Thus, mtts in OI derivation mode (call-by-name) have ``much more''
nondeterminism than mtts in IO derivation mode (call-by-value).
This difference
suggests that translation membership is computationally harder
for OI-mtts than for IO-mtts.

In this paper, we first show that for OI-mtts,
translation membership is NP-complete,
and so is for compositions of multiple IO-mtts (Section~\ref{sec:npcomplete}).
We then present our main result: translation membership for IO-mtts
is solvable in polynomial time (Section~\ref{sec:ptime}).
Our algorithm for IO translation membership
is based on a technique called inverse type inference.
For an mtt $M$ and a given output type, i.e., a regular tree language $L$
of output trees, inverse type inference constructs a description of the
corresponding input type, i.e., of the regular tree language $M^{-1}(L)$.
Note that, inverse type inference basically takes
exponential time, because the size of
the inverse-type automaton itself can be that large~\cite{MSV03,MPS07,PS04}.
To avoid this, we construct the automaton on-the-fly and obtain
the PTIME efficiency.
Our technique is then generalized to several extension of IO-mtts,
such as addition of regular
look-ahead or the generalization to multi-return mtts.
In fact, we even consider a more powerful look-ahead mechanism that
is based on tree automata with equality and disequality 
constraints between siblings~\cite{BT92}.

Note that, for total deterministic mtts: OI equals IO, and
by Theorem~15 of~\cite{Maneth02}, given an input tree $s$,
the output tree $\tau(s)$ can be
computed in time $O(\abs{s}+\abs{t})$,
even for an $n$-fold composition of total deterministic mtts.
%which form a proper hierarchy of translations with respect to $n$
%(see e.g., Theorem 4.16 of~\cite{EV85}).
Hence, by simply computing the output,
translation membership can be solved in linear time for this class
of translations.
The result can easily be extended to deterministic but partial mtts
 (in either IO or OI derivation mode),
 as mentioned at the end of Section~\ref{sec:ptime}.

%%%%%%%%%%%%%%%%%%%%%%%%%%%%%%%%%%%%%%%%%%%%%%%%%%%%%%%%%%%%%%%%%%%%%%%%%%%%%%%%

\section{Definitions}
\label{sec:definitions}

For a finite set $A$, we denote by $\abs{A}$ the number of its elements.
A finite set $\Sigma$ with a mapping
 $\rankfn: \Sigma \to \nat$ is called a {\em ranked alphabet}.
We often write $\sigma^{(k)}$ to indicate that $\rankfn(\sigma) = k$
%and say that $\sigma$ is a rank-$k$ symbol;
and write $\Sigma^{(k)}$ to denote the subset of $\Sigma$ of rank-$k$ symbols.
The {\em product} of $\Sigma$ and a set $B$ is the ranked alphabet
$\Sigma \times B =
  \{\tup{\sigma,b}^{(k)} \mid \sigma^{(k)} \in \Sigma, b \in B \}$.
Throughout the paper, we fix the sets of input variables
$X=\{x_1, x_2, \dots\}$, parameters
$Y=\{y_1, y_2, \dots\}$, and let-variables
$Z=\{z_1, z_2, \dots\}$, which are all of rank 0.
We assume any other alphabet to be disjoint with $X$, $Y$, and $Z$.
The set $X_i$ is defined as $\{x_1, \dots, x_i\}$,
and $Y_i$ and $Z_i$ are defined similarly.

The set $T_\Sigma$ of {\em trees} $t$ over a ranked alphabet $\Sigma$
is defined by the BNF $t \text{\,::=\,} \sigma (\overbrace{t, \dots, t}^{k})$
for $\sigma \in \Sigma^{(k)}$.
We often omit parentheses for rank-0 and rank-1 symbols.
We recursively define the function $\labelfn$ from
$T_\Sigma \times \nat^*$ to $\Sigma$ as follows.
For $t = \sigma(t_1, \ldots, t_k)$, $\sigma^{(k)} \in \Sigma$, $k \geq 0$,
and $t_1, \ldots, t_k \in T_\Sigma$,
  $\labelfn(t, \epsilon) = \sigma$ and
  $\labelfn(t, i . \nu)  = \labelfn(t_i, \nu)$.
Thus, the empty list $\epsilon$ denotes the root node and $\nu.i$
 denotes the $i$-th child of $\nu$.
We define the set
 $\posfn(t) = \{\nu \in \nat^* \mid \labelfn(t,\nu) \text{ is defined}\}$.
We denote by $\abs{t}$ the number of nodes in the tree $t$. 
For a node $v$ of $t$, 
 $t|_v$ denotes the subtree of $t$ rooted at the node $v$.
For trees $t, t_1, \dots, t_n \in T_\Sigma$ and
 $\sigma_1, \dots, \sigma_n \in \Sigma^{(0)}$, 
we denote by $t\,[\sigma_1/t_1, \dots, \sigma_n/t_n]$
 the simultaneous substitution of the $\sigma_i$ by the $t_i$.

Let $\Sigma$ and $\Delta$ be ranked alphabets.
A relation $\tau \subseteq T_\Sigma \times T_\Delta$
 is called a {\em tree translation}
(over $\Sigma$ and $\Delta$) or simply a translation.
We define ${\it range}(\tau) = \{b \mid \exists a: (a,b)\in \tau \}$.
For two translations $\tau_1$ and $\tau_2$,
their sequential composition
 $\tau_1 \comp \tau_2$ (``$\tau_1$ followed by $\tau_2$'')
is the translation
 $\{(a,c) \mid \exists b: ((a,b) \in \tau_1, (b,c) \in \tau_2) \}$.
For two classes $T_1$ and $T_2$ of translations, we define
 $T_1 \comp T_2 =
   \{\tau_1 \comp \tau_2 \mid \tau_1 \in T_1, \tau_2 \in T_2 \}$.
The $k$-fold composition of the class $T$ of translations
 is denoted by $T^k$.

\begin{definition}\label{def_mtt}
A {\em macro tree transducer (mtt)} $M$ 
 is a tuple
    $(Q,\Sigma,\Delta,q_0,R)$,
 where
   $Q$ is the ranked alphabet of {\em states},
   $\Sigma$ and $\Delta$ are the {\em input} and {\em output} alphabets,
   $q_0 \in Q^{(0)}$ is the {\em initial state}, and
   $R$ is the finite set of {\em rules} of the form
   \[
      \tup{q, \sigma(x_1,\dots,x_k)} (y_1, \ldots, y_m) \to r
   \]
where $q \in Q^{(m)}$, $\sigma \in \Sigma^{(k)}$,
 and $r$ is a tree in $T_{\Delta \cup (Q \times X_k) \cup Y_m}$.
Rules of such form are called $\tup{q,\sigma}$-rules,
 and the set of right-hand sides of
all $\tup{q,\sigma}$-rules is denoted by $R_{q,\sigma}$.
We define the size of the mtt by
 $\abs{M} = \sum \{\abs{r} \mid r \in R_{q,\sigma}, q\in Q, \sigma\in\Sigma \}$.
\end{definition}

For the remainder of this section, let $M$ be an mtt as in
Definition~\ref{def_mtt}.
A state $q$ of a macro tree transducer
 can be regarded as a (nondeterministic) {\em function}
in functional programming languages.
Depending on the order of evaluation,
two different semantics can be considered: call-by-value (or inside-out, IO)
and call-by-name (or, outside-in, OI).
Let $\mu \in \{\IO, \OI\}$.
For the tree $u \in T_{\Delta\cup (Q \times T_\Sigma)\cup Y}$,
its meaning with respect to $M$
$\sem{u}^M_\mu \subseteq T_{\Delta\cup Y}$ is inductively defined as follows
\begin{align*}
  \sem{ y_i }^M_\mu
    &=
       \{ y_i \}
\\
  \sem{ \delta(u_1, \ldots, u_n) }^M_\mu
    &=
      \{ \delta(t_1, \ldots, t_n) \mid
      \\\tag*{$\displaystyle %%%% Right-align
        t_i \in \sem{u_i}^M_\mu \text{ for all } i\}
      $} %%%% Right-align
\\
  \sem{ \tup{q,\sigma(s_1, \ldots, s_k)}(u_1, \ldots, u_m) }^M_\mu
    &=
      \\\tag*{$\displaystyle %%%% Right-align
        \bigcup_{r \in R_{q,\sigma}} \Big(
          \bigsem{ r[x_1/s_1, \dots, x_k/s_k] }^M_\mu
                \xleftarrow[\mu]{} (\sem{u_1}^M_\mu, \ldots, \sem{ u_m }^M_\mu)
        \Big)
      $} %%%% Right-align
\end{align*}
where $\xleftarrow[\rm IO]{}$ and $\xleftarrow[\rm OI]{}$
denote IO- and OI-substitution, respectively, and are defined
as follows for $L, L_1,\dots, L_n\subseteq T_{\Delta \cup Y}$.
\begin{align*}
  L \xleftarrow[\mu]{} (L_1, \dots, L_n)
    &=
      \bigcup_{t \in L} \left( t \xleftarrow[\mu]{} (L_1, \dots, L_n) \right)
\\
  t \xleftarrow[\rm IO]{} (L_1, \dots, L_n)
    &=
      \{t[y_1/t_1, \dots, y_n/t_n] \mid
      \\\tag*{$\displaystyle %%%% Right-align
        t_i \in L_i \text{ for all } i\}
      $} %%%% Right-align
\\
  y_i \xleftarrow[\rm OI]{} (L_1, \dots, L_n)
    &=
      L_i
\\
  \delta(t_1, \dots, t_m) \xleftarrow[\rm OI]{} (L_1, \dots, L_n)
    &=
      \\\tag*{$\displaystyle %%%% Right-align
        \{ \delta(t'_1, \dots, t'_m) \mid
           t'_i \in
           \left( t_i \xleftarrow[\rm OI]{} (L_1, \dots, L_n)\right) \text{ for all } i
        \}.
      $} %%%% Right-align
\end{align*}
The difference of IO- and OI- semantics lies in the interpretation of state calls.
In IO-semantics we use IO-substitution for parameters; each parameter $y_i$ is bound
to some fixed (but nondeterministically chosen) tree in $\sem{u_i}^M_\IO$, and
every occurrence of $y_i$ is replaced with the same single tree.
On the other hand, in OI-semantics,
each parameter is bound to the set of trees $\sem{u_i}^M_\OI$, and
at every occurrence of $y_i$ we nondeterministically choose some tree
in $\sem{u_i}^M_\OI$, independent from the choices made at other occurrences of $y_i$.

As an example of the definition of $\sem{u}_\mu$,
consider the example from the Introduction.
Note that there we used slightly different notation:
the right-hand side $\op{double}(x, \op{double}(x, {\tt e}))$
is now written as
$\tup{\op{double}, x_1}(\tup{\op{double}, x_1}({\tt e}))$,
i.e., we distinguish the first parameter---which is the special parameter
that is bound to an input tree in $T_\Sigma$---from others bound to output trees in $T_\Delta$,
by enclosing it with angle brackets.
Now, let us compute $\sem{\tup{\op{start},{\tt a}({\tt a}({\tt e}))}}_\mu$.
\begin{align*}
  &\sem{\tup{\op{start},{\tt a}({\tt a}({\tt e}))}}_\mu
\\
=\ &
 \sem{\tup{\op{double},{\tt a}({\tt e})}( \tup{\op{double},{\tt a}({\tt e})}({\tt e}) )}_\mu
\\
=\ &
  \sem{\tup{\op{double},e}( \tup{\op{double},{\tt e}}(y_1) )}_\mu
\xleftarrow[\mu]{}
  \sem{\tup{\op{double},{\tt a}({\tt e})}({\tt e})}_\mu
\\[-0.6em]
=\ &
\Big(
    \{{\tt f}(y_1,y_1), {\tt g}(y_1,y_1)\}
  \xleftarrow[\mu]{}
    \sem{\tup{\op{double},{\tt e}}({\tt e})}_\mu
\Big)
      \\[-0.5em]\tag*{$\displaystyle %%%% Right-align
\xleftarrow[\mu]{}
  \sem{\tup{\op{double},{\tt a}({\tt e})}({\tt e})}_\mu
      $} %%%% Right-align
\\[-0.6em]
=\ &
\Big(
    \{{\tt f}(y_1,y_1), {\tt g}(y_1,y_1)\}
  \xleftarrow[\mu]{}
    \{{\tt f}(y_1,y_1), {\tt g}(y_1,y_1)\}
\Big)
      \\[-0.5em]\tag*{$\displaystyle %%%% Right-align
\xleftarrow[\mu]{}
  \sem{\tup{\op{double},{\tt a}({\tt e})}({\tt e})}_\mu
      $} %%%% Right-align
\end{align*}
Here, we encountered the $\mu$-substitution
$L\xleftarrow[\mu]{} L$ for $L=\{{\tt f}(y_1,y_1), {\tt g}(y_1,y_1)\}$.
Now, if $\mu=\IO$ then 
$L \xleftarrow[\mu]{} L = \{
{\tt f}({\tt f}(y_1,\break y_1),{\tt f}(y_1,y_1)),\ 
{\tt g}({\tt f}(y_1,y_1),{\tt f}(y_1,y_1)),\ 
{\tt f}({\tt g}(y_1,y_1),{\tt g}(y_1,y_1)),\break
{\tt g}({\tt g}(y_1,y_1),{\tt g}(y_1,y_1))\}$;
the size of the set is $2 \times 2 = 4$.
On the other hand, if $\mu=\OI$ then we obtain
$L\xleftarrow[\mu]{} L=\{
{\tt f}({\tt f}(y_1,y_1),{\tt f}(y_1,y_1)),\ \ 
{\tt f}({\tt f}(y_1,y_1),{\tt g}(y_1,y_1)),\ \ 
{\tt f}({\tt g}(y_1,y_1),\break {\tt f}(y_1,y_1)),\ \ \ 
{\tt f}({\tt g}(y_1,y_1),{\tt g}(y_1,y_1)),\ \ \ 
{\tt g}({\tt f}(y_1,y_1),{\tt f}(y_1,y_1)),\break
{\tt g}({\tt f}(y_1,y_1),{\tt g}(y_1,y_1)),\ \ 
{\tt g}({\tt g}(y_1,y_1),{\tt f}(y_1,y_1)),\ \ 
{\tt g}({\tt g}(y_1,y_1),\break {\tt g}(y_1,y_1))\}$;
the size is $2 \times 2^2 = 8$ where the exponent $2$ comes from the
number of occurrences of the parameter $y_1$ in each target term of the substitution.

We define
the {\em translation realized by} $M$ in $\mu$-mode by the relation
$\tau_{\mu,M} = \{(s,t) \in T_\Sigma \times T_\Delta \mid t \in \sem{\tup{q_0,s}}_\mu \}$.
The class of all translations realized by all mtts in $\mu$-mode is denoted by $\MTT_\mu$.
An mtt is called {\em deterministic} (respectively, {\em total\/}) if for every $q, \sigma$,
the number of rules $\abs{R_{q,\sigma}}$ is at most (at least) 1;
the corresponding classes of translations are  denoted by prefix $\text{\rm D}$ (${\rm t}$).
An mtt is called {\em linear (in the input variables)} if in every right-hand side of the rules,
each input variable $x_i$ appears at most once; the corresponding class of translation
is denoted by prefix $\text{L}$.
For example, the class of translations realized by linear, deterministic, and total mtts in OI mode
is denoted by ${\rm LD}{\rm t}{\rm MTT}_\OI$.

For a translation $\tau \subseteq T_\Sigma \times T_\Delta$,
the {\em translation membership problem} for $\tau$ is a decision problem
that determines, given a tree $s \in T_\Sigma$ and a tree $t \in T_\Delta$,
whether $(s,t) \in \tau$.
In the rest of the paper, we focus on the {\em data complexity} of this problem.
That is, we measure the complexity in terms of $\abs{s} + \abs{t}$,
regarding the translation $\tau$ to be fixed.
We will always assume that the input and output tree that are
inputs to the problem are denoted by ``$s$'' and ``$t$''.

%%%%%%%%%%%%%%%%%%%%%%%%%%%%%%%%%%%%%%%%%%%%%%%%%%%%%%%%%%%%%%%%%%%%%%%%%%%%%%%%

\section{NP-complete Classes}
\label{sec:npcomplete}

The first result is that translation membership for OI-mtts is
NP-hard, even for linear mtts. The proof is based on the reduction to 3-SAT,
which resembles~\cite{Rounds73} which shows NP-completeness of the membership
problem for indexed languages.
In fact, the indexed languages can be obtained as yields (strings of leaves from
left to right) of output languages of linear mtts (by the fact that each indexed language
is the yield of some OI context-free tree language~\cite{Fischer68} and
each OI context-free tree language is equivalent to the range
${\it range}(\tau)$ of some $\tau \in \LMTTOI$ by Corollary 6.13 in~\cite{EV85}).
However, given a word $w$ as input for the membership problem of an indexed language
$L$, it is not clear how to construct a pair $(s,t)$ such that $(s,t) \in \tau\/_{\OI,M}$
for some linear mtt if and only if $w$ is in $L$.
We can choose $s=a^n$ with $n={\it length}(w)$ and an $\LMTT$
which produces trees $t$ which have as yield the word $w$.
But how to select such a tree $t$ as input for the translation membership problem?
Note that it is easy to construct from $w$ an input for translation
membership for a two-fold composition of mtts: the second transducer
realizes ``yield'', i.e., it turns a tree $t$ into a monadic tree
that represents $t$'s yield (such a transducer is even total deterministic).
Thus, it follows that translation membership
for two-fold compositions of mtts is NP-hard. This was mentioned already
in~\cite{IM08}. The next lemma shows that even translation
membership for a single linear mtt is NP-hard.

\begin{lemma}\label{LMTTOI}
Translation membership for $\LMTTOI$ \hspace{-0.9pt}(\hspace{-0.5pt}and hence $\MTTOI$) is NP-hard.
\end{lemma}
\begin{proof}
We construct an mtt $M=(Q,q_0,\Sigma,\Delta,R)$ so that it generates the parse-trees
 of all satisfiable boolean formulas in 3-conjunctive normal form, given the number of
 variables $n$ and clauses $m$ as the inputs.
We slightly abuse our notation and write $y_v,y_t,y_f$ in place of $y_1,y_2,y_3$, respectively.
Let $Q=\{q_0^{(0)}, q_c^{(2)}, \allowbreak q^{(3)}\}$,
$\Sigma=\{\syma^{(1)},\symb^{(3)},\symc^{(1)},\symd^{(0)}\}$,
$\Delta=\{\wedge^{(2)}, \vee^{(3)}, \neg^{(1)}, \symv^{(1)}, \allowbreak \syme^{(0)}\}$,
and $R$ the following set of rules:
\begin{align*}
   \tup{q_0, \syma(x_1)}
      & \to
   \tup{q, x_1}(\symv(\syme), \syme, \neg(\syme))
\\
   \tup{q_0, \syma(x_1)}
      & \to
   \tup{q, x_1}(\symv(\syme), \neg(\syme), \syme)
\\
   \tup{q, \symb(x_1,x_2,x_3)} (y_v, y_t, y_f)
      & \to \\
   \tup{q, x_1}(  \symv(y_v), \tup{q_c,x_2}&(y_t, y_v), \tup{q_c,x_3}(y_f, \neg(y_v)) )
\\
   \tup{q, \symb(x_1,x_2,x_3)} (y_v, y_t, y_f)
      & \to \\
   \tup{q, x_1}(  \symv(y_v), \tup{q_c,x_2}&(y_t, \neg(y_v)), \tup{q_c,x_3}(y_f, y_v) )
\\
   \tup{q_c, \symd}(y_1, y_2)
      & \to y_1
\\
   \tup{q_c, \symd}(y_1, y_2)
      & \to y_2
\end{align*}
\begin{align*}
   \tup{q, \symc(x_1)}(y_v, y_t, y_f)
      & \to \wedge( \vee(y_t, y_t, y_t), \tup{q, x_1}( y_v, y_t, y_f ) )
\\
   \tup{q, \symc(x_1)}(y_v, y_t, y_f)
      & \to \wedge( \vee(y_t, y_t, y_f), \tup{q, x_1}( y_v, y_t, y_f ) )
\\
   \tup{q, \symc(x_1)}(y_v, y_t, y_f)
      & \to \wedge( \vee(y_t, y_f, y_t), \tup{q, x_1}( y_v, y_t, y_f ) )
\\
   \tup{q, \symc(x_1)}(y_v, y_t, y_f)
      & \to \wedge( \vee(y_f, y_t, y_t), \tup{q, x_1}( y_v, y_t, y_f ) )
\\
   \tup{q, \symc(x_1)}(y_v, y_t, y_f)
      & \to \wedge( \vee(y_t, y_f, y_f), \tup{q, x_1}( y_v, y_t, y_f ) )
\\
   \tup{q, \symc(x_1)}(y_v, y_t, y_f)
      & \to \wedge( \vee(y_f, y_f, y_t), \tup{q, x_1}( y_v, y_t, y_f ) )
\\
   \tup{q, \symc(x_1)}(y_v, y_t, y_f)
      & \to \wedge( \vee(y_f, y_t, y_f), \tup{q, x_1}( y_v, y_t, y_f ) )
\\
   \tup{q, \symd}(y_v, y_t, y_f)
      & \to \vee(y_t, y_t, y_f)
\\
   \vdots \quad \text{(same}&\text{ as the $\vee(\cdots)$ part of $\tup{q,c}$-rules)}
\\
   \tup{q, \symd}(y_v, y_t, y_f)
      & \to \vee(y_f, y_t, y_f).
\end{align*}
From an input tree $\syma(\overbrace{\symb(\symb(\cdots \symb(}^n\symc^m\symd, \symd, \symd)\cdots), \symd, \symd))$
of size $3n + m + 2$, it generates all satisfiable boolean formulas in 3-conjunctive normal form
with $n$ variables and $m$ conjuncts.
The output language encodes boolean formulas as follows:
a boolean variable $p_i$ for $0 \leq i < n$ is represented as $\symv^i\syme$, 
and three boolean operations $\neg$, $\wedge$, and $\vee$ are represented as they are.
For example, the formula $(p_0 \vee \neg p_1 \vee p_2) \wedge (\neg p_0 \vee p_1 \vee p_2)$
is encoded as $\wedge(\vee(\syme, \neg \symv\syme, \symv\symv\syme), \vee(\neg \syme, \symv\syme, \symv\symv\syme))$.

Intuitively, when the mtt reads the root node of the input,
 it nondeterministically assigns a truth-value to the first variable $p_0$.
The first $\tup{q_0,\syma}$-rule is the case when it assigned `true' and the other rule is for `false'.
Three parameters are passed to the state $q$. Intuitively, the first parameter $y_v$
denotes the name of the next variable to be assigned a truth-value.
The second (and the third, respectively) parameter $y_t$ ($y_f$) denotes
the set of `true' (`false') literals (namely, variables or negated variables)
that have been constructed up to now.
While reading $\symb$ nodes in the state $q$, the mtt nondeterministically assigns
a truth-value to each variable $p_1$ to $p_{n-1}$, similarly to $p_0$.
Here, OI-nondeterminism is crucially used to represent
arbitrary choice of positive and negative literals; each time $y_t$ and $y_f$
are copied to the output, they contain unevaluated ``combs'' of
$q_c$-calls (on $\symd$-nodes). Each such comb represents the nondeterministic choice 
of any of the positive ($y_t$) or negative ($y_f$) literals that have been
generated so far.
The state $q_c$ means a union of two sets,
by taking two parameters and nondeterministically returns either one of them.
The parameter $y_t$ is assigned an unevaluated expression, e.g., like
$\tup{q_c,d}(\tup{q_c,d}(\neg p_0, p_1),p_2)$,
and each time the value of $y_t$ is needed, it is nondeterministically
evaluated to either $\neg p_0$, $p_1$, or $p_2$.
Then, while reading $\symc$ nodes in the input, the transducer generates $m$ conjunctions of
`true' clauses. Since we generate 3-CNF formulas,
each clause consists of a disjunction of exactly three literals.
There are seven possibilities (all combinations of $y_t$ and $y_f$, except
$\vee(y_f, y_f, y_f)$), which are generated by the $\tup{q,\symc}$-rules of the transducer.

It should be clear for the reader that this mtt generates all (and only) satisfiable 3-CNF
formulas; it nondeterministically constructs any of the $2^n$ possible assignments
to the variables $p_0, \dots, p_{n-1}$,
 and under each assignment, generates any of the possible $7^m$ types of
`true' formulas.
The point is, the choices at $\tup{q_c,\symd}$ for enumerating all possible literals
are nondeterministically evaluated each time generating a disjunct,
while the choices at $\tup{q_0,\syma}$ and $\tup{q,\symb}$ for enumerating all possible
truth-value assignments are
evaluated and uniformly determined prior to the generation of all conjuncts.

It is also obvious that, given any 3-CNF formula, we can in polynomial time
encode the formula to the above explained encoding to obtain $t$, and count the number of
variables and clauses to obtain $s$. Then, $(s,t) \in \tau_M$ if and only if the original
formula is satisfiable. It is well known that the satisfiability of 3-CNF is NP-complete
(see, e.g., \cite{GareyJohnson}).
\end{proof}

In \cite{IM08}, we have proved two closely related results;
one is that the above NP-hard lowerbound is tight, i.e., the translation membership
for $\LMTTOI$ can be determined in NP time complexity. The other is
that the complexity of membership problem of the {\em output language} is in NP,
even for finitely many compositions of $\MTTOI$'s.
Altogether, we have the following theorem.

\begin{theorem}\label{MTTOIn}
Translation membership for $\MTTOI^n$ for $n \geq 1$ is NP-complete.
\end{theorem}
\begin{proof}
NP-hardness follows from the preceding lemma.
Let $\tau \in \MTTOI^n$.
We can easily construct a translation $\tau' = \{(s,\pi(s,t)) \mid (s,t) \in \tau\}$
in $\MTTOI^n$ where $\pi$ is a new binary symbol.
This is done by changing the first mtt $M_1$ (with input alphabet $\Sigma$ and
initial state $q_0$) of the composition as follows.
Replace for $\sigma \in \Sigma^{(k)}$ every $\tup{q_0, \sigma}$-rule
with right-hand side $t$ by the new rule
$\tup{q_0, \sigma(x_1, \dots, x_k)} \to \pi(
 \sigma(\tup{q_{\it id},x_1}, \dots, \tup{q_{\it id},x_k})
,t)$ and introduce
$\langle q_{\it id}, \sigma(x_1, \dots, \allowbreak x_k) \rangle \to
 \sigma(\tup{q_{\it id},x_1}, \dots, \tup{q_{\it id},x_k})$
for the new state $q_{\it id}$ of rank 0.
Then, the subsequent mtts $M_i$ ($2 \leq i \leq n$) are augmented by the new rule
$\tup{q_0, \pi(x_1,x_2)} \to \pi(\tup{q_{\it id}, x_1},\tup{q_0,x_2})$
and $q_{\it id}$ rules as for $M_1$.
Note that $(s,t) \in \tau$ if and only if $\pi(s,t) \in {\it range}(\tau')$.
Since by Theorem 8 of~\cite{IM08} the complexity of the membership test of ${\it range}(\tau')$
is in NP, we can also check $(s,t) \in \tau$ in NP.
\end{proof}

Note that compositions of two $\MTTIO$'s can simulate all $\MTTOI$
translations (Theorem 6.10 of~\cite{EV85}), and conversely, compositions
of $\MTTIO$'s can be simulated by compositions $\MTTOI$'s (Theorem 7.8 of~\cite{EV85}). Therefore,
we now have the NP-completeness for compositions of $\MTTIO$'s.
\begin{corollary}\label{MTTIOn}
Translation membership for $\MTTIO^n$ for $n \geq 2$ is NP-complete.
\end{corollary}

%%%%%%%%%%%%%%%%%%%%%%%%%%%%%%%%%%%%%%%%%%%%%%%%%%%%%%%%%%%%%%%%%%%%%%%%%%%%%%%%

\section{Tractable Classes}
\label{sec:ptime}

In this section, we first prove that IO-mtts have polynomial-time
translation membership, contrary to OI-mtts.
Then we extend the result to several other extensions of IO-mtts,
and to some restricted subclasses of OI-mtts.

The idea of the proof is based on inverse type inference for mtts M
(Theorem 7.4 of \cite{EV85});
given a finite tree automaton $\mathcal{B}$ (accepting output trees),
we can effectively construct a finite tree automaton that
recognizes the corresponding input trees $\tau^{-1}_M( L(\mathcal{B}) )$.
Given an output tree $t$, by constructing its minimal dag representation
(i.e., the pointer representation of $t$ such that all isomorphic subtrees are shared),
we can simply consider it as the trivial deterministic automaton $\mathcal{B}_t$
with at most $\abs{t}$-many states which recognizes $\{t\}$.
Once we have constructed the automaton $\mathcal{A}$ for $\tau^{-1}_M( L(\mathcal{B}_t) )$,
we merely need to check whether $s \in L(\mathcal{A})$, in order to solve
translation membership for $(s,t)$.
However, the automaton $\mathcal{A}$ can be very large: its worst case number of
states is exponential in $\abs{\mathcal{B}_t}$.
Thus, we must avoid to fully construct $\mathcal{A}$ in order to obtain PTIME complexity.
Our idea is to construct $\mathcal{A}$ on demand, while running it on the tree $s$.
Note that inverse type inference of an IO-mtt constructs an input type automaton
which has states that are functions $p$ from $Q$ to $(V^m \to 2^V)$ where $V$
is the set of states of $\mathcal{B}_t$, $Q$ is the set of states of $M$,
and $m$ is the maximum rank of states in $Q$.
Such a state $p$ tells us for each $q \in Q$, which state of $\mathcal{B}_t$
is obtained if we apply the state $q$ to an input tree.
That is, if $\mathcal{A}$ reaches the state $p$ after reading a tree $s$,
it means that running $\mathcal{B}_t$ on output trees in
$\tup{q,s}(t|_{v_1}, \ldots, t|_{v_m})$ obtains the states $(p(q))(v_1, \dots, v_m)$.
%
%Such a state $p$ tells us for each $q \in Q$, which state of $\mathcal{B}_t$
%is obtained, if we know that for $q$'s parameter trees, $\mathcal{B}_t$ arrives
%in states $v_1, \dots, v_m$ of $\mathcal{B}_t$:
%it is the state $(p(q))(v_1, \dots, v_m)$.

%%%%%%%%%%%%%%%%%%%%%%%%%%%%%%%%%%%%%%%%%%%%%%%%%%%%%%%%%%%%%%%%%%%%%%%%%%%%%%%%

\begin{theorem} \label{MTTIO}
Let $M$ be an mtt.
\hspace{-1.95pt}Translation membership for $\tau\/_{\IO,M}$ can be determined
in time $O(\abs{s} \cdot \abs{t}^{2m+2} \cdot \abs{M})$
where $m$ is the maximum rank of $M$'s states.
\end{theorem}

\begin{proof}
Let $t_{\it dag}$ be the minimal dag representing $t$.
It is folklore that $t_{\it dag}$ can be computed in 
amortized linear time in $\abs{t}$, using hashing,
and even in linear time using pseudo radix sorting, see~\cite{DST80}.
Let $V_t$ be the set of nodes of $t_{\it dag}$.
We define $\labelfn(v)$ to denote the label in $\Sigma$ of the node $v \in V_t$,
and $\childfn(v, i)$ to denote the $i$-th child node of $v$.
Assuming a standard pointer structure representing dags,
we regard each execution of $\labelfn$ and $\childfn$ takes $O(1)$ time.

Let $\bot$ be an element distinct from $V_t$.
Let $V = V_t \cup \{\bot\}$ and $\labelfn(\bot)$ to be undefined.
Let ${\it run} : T_\Sigma \to A$ with $A = 2^{\bigcup_i Q^{(i)} \times V^i \times V}$
 be the function defined inductively as follows
\[
  {\it run}( \sigma(s_1, \dots, s_k) ) = {\it tr}(
    \sigma, {\it run}(s_1), \dots, {\it run}(s_k)
  )
\]
where ${\it tr}$ is defined below.
The set $A$ contains the states of the deterministic bottom-up automaton of
$\tau^{-1}(t)$, ${\it tr}$ is the transition function, and ${\it run}$ computes the
run of the automaton.
The intuition of the set of states $A$ is, that
``$(q,\vec{v},v') \in {\it run}(s')$'' means that ``if $q$ is applied to
the input subtree $s'$ with output subtrees rooted at $\vec{v}$ as parameters,
then it may generate an output subtree rooted at $v'$''.
The special value $\bot \in V$ is used to denote a tree that is not a subtree of $t$.
That is, for example, ``$(q,\vec{v},\bot) \in {\it run}(s')$'' means that an application
of $q$ to $s'$ with parameters $\vec{v}$ may yield a tree that is not
a subtree of $t$. 

The transition function ${\it tr} : (\bigcup_i \Sigma^{(i)} \times A^i) \to A$ is defined as follows
\[
  {\it tr}( \sigma, \vec{a} ) =
   \Big\{
      (q, \vec{v}, v') \in \bigcup_i Q^{(i)} \times V^i \times V
      \\\tag*{$\displaystyle %%%% Right-align
    \Big|\ 
        \exists r \in R_{q,\sigma}:
          f_{\vec{v}, \vec{a}}(r, v')
    \Big\}
      $} %%%% Right-align
\]
where $f_{\vec{v}, \vec{a}}
 : T_{\Delta \cup (Q \times X) \cup Y} \times V \to \{{\it true}, {\it false}\}$
is defined inductively on right-hand sides of the rules:
\begin{align*}
f_{\vec{v}, \vec{a}}(y_i, v') & = {\it true}
  \hspace{67pt}
  \text{if } v' = v_i
\\
f_{\vec{v}, \vec{a}}(y_i, v') & = {\it false}
  \hspace{65pt}
  \text{if } v' \neq v_i
\\
f_{\vec{v}, \vec{a}}(\delta(r_1, \dots, r_n), v') & =\\
  \labelfn(v') =
 \delta \wedge
  &\hspace{-6pt}
  \bigwedge_{1 \leq i \leq n} \hspace{-6pt} f_{\vec{v}, \vec{a}}(r_i, {\it child}(v',i))
  \hspace{7.5pt}
  \text{if } v' \in V_t
\\
f_{\vec{v}, \vec{a}}(\delta(r_1, \dots, r_n), \bot) & =
  \big(\exists \vec{u} \in V^n:
    \hspace{-6pt}
      \bigwedge_{1 \leq i \leq n} \hspace{-6pt} f_{\vec{v}, \vec{a}}(r_i, u_i)
  \big)
    \wedge
\\
  \big(
    \forall u' \in V_t: \neg\big(
      \labelfn(&u') = \delta
      \wedge
      \bigwedge_{1 \leq i \leq n} \hspace{-6pt} \childfn(u',i) \neq u_i
    \big)\big)
\\
f_{\vec{v}, \vec{a}}(\tup{q', x_j}(r_1, \dots, r_n), v') & =
\\
  \exists \vec{u} \in V^n:
   \Big(
     (q', &\vec{u}, v') \in a_j \wedge
         \bigwedge_{1 \leq i \leq n} f_{\vec{v}, \vec{a}}(r_i, u_i)
   \Big).
\end{align*}
The relation $f_{\vec{v}, \vec{a}}(r, v')$ should be understood as:
``evaluation of $r$ will yield the output subtree at $v'$, under the assumption that
  the parameters $\vec{y}$ are bound to $\vec{v}$ and
  the effects of application of a state to each child is as described by $\vec{a}$\,''.

For a tree $t' \in T_\Delta$, let $\rho(t')$ be $v \in V_t$ if $t' = t|_v$,
and $\rho(t') = \bot$ otherwise.
We also define $\rho(T)$ for $T \subseteq T_\Delta$ as $\{\rho(t) \mid t \in T\}$.
The correctness of the above construction is verified by
the following claim. Note that the claim is just rephrasing the intuition of the set of states $A$
explained above, in a formal way.

\textit{Claim}\quad 
For every input tree $s'$, we have the following equation for all
 $q \in Q$, 
 $r_i \in T_{\Delta \cup (Q \times T_\Sigma)}$, and an environment $\Gamma$:
$
\rho\Big( \sem{\tup{q,s'}(r_1,\dots,r_n)}^M_\IO \Big)
=
 \Big\{
   v' \ \Big|\  (q,(v_1,\ldots,v_n),v') \in {\it run}(s'), v_i \in \rho(\sem{r_i}^M_\IO)
   \text{ for all }i
 \Big\}
$

By applying the claim for $q=q_0$ and $s'=s$,
we know that $t \in \sem{\tup{q,s}}^M_\IO$
is equal to  $(q_0, (), v_\epsilon) \in {\it run}(s)$
where $v_\epsilon$ is the root node of $t_{\it dag}$.
Hence, the translation membership can be determined by computing the set ${\it run}(s)$.

The proof of the claim is by nested induction first on structure of
 $s'$, and then on the structure of right-hand sides of the rules.
Let $s' = \sigma(s_1, \dots, s_k)$ (the base case is the case $k=0$).
By definition of the IO-semantics we have
\begin{align*}
\rho\Big( \sem{\tup{q,s'}(r_1,\dots,r_n)}^M_\IO \Big)
= \bigcup_{r \in R_{q,\sigma}} \hspace{-6pt}\Big\{
   \rho(t'[y_1/t_1, \ldots, y_n/t_n])
  \ \Big|\
      \\\tag*{$\displaystyle %%%% Right-align
    t' \in \sem{r[\vec{x}/\vec{s}]}^M_\IO,
    t_i \in \sem{r_i}^M_\IO \text{ for all } i
  \Big\}
      $} %%%% Right-align
\end{align*}
and by definition of ${\it run}$, we have
\begin{align*}
 \Big\{
   v' \ \Big|\  (q,\vec{v},v') \in {\it run}(s'), v_i &\in \rho(\sem{r_i}^M_\IO)
 \Big\}
\\
=
 \bigcup_{r \in R_{q,\sigma}}
 \Big\{
   v'
   &\ \Big|\ 
   f_{\vec{v},\vec{a}}(r, v'), v_i \in \rho(\sem{r_i}^M_\IO)
 \Big\}
\end{align*}
where $\vec{a} = ({\it run}(s_1), \ldots, {\it run}(s_k))$.
To show these two sets are equal, it is sufficient to prove the
the following statement:
 if $\rho(t_i) = v_i$ then
$\{\rho(t'[\vec{y}/\vec{t}]) \mid t' \in \sem{r[\vec{x}/\vec{s}]}^M_\IO \}
= \{v' \mid f_{\vec{v},\vec{a}}(r, v')\}$.
The proof is by nested induction on the structure of $r$.
For example, if $r = \tup{q',x_i}(r_1, \ldots, r_n)$,
we have
 $\{v' \!\mid f_{\vec{v},\vec{a}}(\tup{q',x_i}(r_1, \ldots, r_n), v')\}
=
 \{v' \mid 
    (q', \vec{u}, v') \allowbreak \in a_i,
      f_{\vec{v}, \vec{a}}(r_i, u_i) \text{ for all } i
 \}$\!,
which is by inner induction hypothesis equal to
 $\{v' \!\mid\!
    (q', \vec{u}, v') \!\in a_i,
      u_i \in \rho(\sem{r_i[\vec{x}/\vec{s},\vec{y}/\vec{t}]}^M_\IO)
      \!\text{ for all }\allowbreak i
 \}$,
 and then by outer induction hypothesis it is equal to
 $\rho( \sem{\tup{q',s_i}(r_1[\vec{x}/\vec{s},\vec{y}/\vec{t}], \ldots,
 r_n[\vec{x}/\vec{s},\vec{y}/\vec{t}])}^M_\IO )
 =\{\rho(t'[\vec{y}/\vec{t}]) \mid t' \allowbreak\in \sem{r[\vec{x}/\vec{s}]}^M_\IO \}$.
The other cases are proved similarly.

The time complexity for testing $(q_0,(),v_\epsilon) \in {\it run}(s)$
is computed as follows.
The value ${\it run}(s)$ for the whole input tree $s$ can be
computed by executing the ${\it tr}$ function on each node of $s$.
The computation is done in bottom-up fashion as bottom-up tree automata does,
so that the states in $\vec{a}$ are already constructed.
The number of execution of the ${\it tr}$ function is $\abs{s}$.
The set ${\it tr}(\sigma, \vec{a})$ can be constructed by simply testing
all combinations of $(q,\vec{v},v') \in \bigcup_i Q^{(i)} \times V^i \times V$
 (which is of size $\leq \abs{Q}\cdot\abs{V}^{m+1}$) and
 $r \in R_{q,\sigma}$ by $f_{\vec{v}, \vec{a}}$.
Note that $f_{\vec{v}, \vec{a}}$ may receive $\abs{r} \cdot \abs{V}$ different pairs of arguments, and
the computation of each value $f_{\vec{v}, \vec{a}}(r', v')$ takes $O(\abs{V}^m)$
 time in the worst case (the $f_{\vec{v}, \vec{a}}(\tup{q',x_j}(\cdots))$ case) 
 assuming the values of $f_{\vec{v}, \vec{a}}$
 are already computed for all subexpressions of $r'$.
Hence, $O(\abs{r}\cdot \abs{V}^{m+1})$ time is sufficient here.
Note that the $f_{\vec{v}, \vec{a}}(\delta(\cdots), \bot)$ case
can be computed efficiently in $O(\abs{V})$ time
by remembering the number $\abs{\{v \mid f_{\vec{v},\vec{a}}(r',v)\}}$
for each sub-expression $r'$: the existence of $\vec{u}$ can be checked by
verifying the number is non-zero,
and the check $\childfn(u', i) \neq u_i$ is replaced with
``either not $f_{\vec{v},\vec{a}}(r',\childfn(u', i))$ or the number is more than one''.
Since it is only required to compute the $f_{\vec{v}, \vec{a}}(\delta(\cdots), \bot)$
cases at most $\abs{r}$ times, the time complexity for the cases is $O(\abs{r}\cdot\abs{V})$,
which is subsumed by $O(\abs{r}\cdot \abs{V}^{m+1})$.
Altogether, multiplying all of them yields the desired complexity bound
$O(\abs{s} \cdot \abs{t}^{2m+2} \cdot \abs{M})$.
Note that we have $\abs{V} \leq \abs{t}+1$ by definition,
and that the parameter $\abs{M}$ subsumes
$\Sigma_{q \in Q, r \in R_{q,\sigma}} \abs{r}$.
\end{proof}

%%%%%%%%%%%%%%%%%%%%%%%%%%%%%%%%%%%%%%%%%%%%%%%%%%%%%%%%%%%%%%%%%%%%%%%%%%%%%%%%

The reader may wonder why the same approach does not work for OI-mtts,
whose inverses also preserve the regular tree languages.
The problem is, for OI, the states of the inferred automata are in
$A = 2^{\bigcup_i Q^{(i)} \times (2^V)^i \times V}$ instead of
$A = 2^{\bigcup_i Q^{(i)} \times V^i \times V}$.
The difference is intuitively explained as follows:
in IO-mtts, every copy of a same parameter is an identical output tree
and thus corresponds to a single node in $V$,
while in OI-mtts, each copy is evaluated independently and thus may
correspond to different output nodes.
To capture this phenomenon in the inverse type inference, each parameter must be represented
by a {\em set of} nodes rather than a {\em single} output node.
The additional exponential implies that a single state in $A$
(a subset of $\bigcup_i Q^{(i)} \times (2^V)^i \times V$) can already be exponentially large.
Therefore, on-the-fly construction
does not help to obtain a PTIME algorithm.
Of course, Lemma~\ref{LMTTOI} implies that there is no PTIME algorithm
for translation membership for OI-mtts (unless NP=P).

Nevertheless, some subclasses of OI-mtts still admit PTIME translation membership.
Note that the essential difficulty of OI-translation membership comes from the copying of parameters.
Consider, for example, an OI-mtt that is linear in the 
parameters (i.e., in every right-hand side each parameter
$y_i$ occurs at most once); then each parameter is either used once or is never used.
In this case, it can be represented in the inverse-type automaton by a set of size $\leq 1$.
More generally, if an OI-mtt is {\em finite copying} in the parameter,
its translation membership can be tested in polynomial time.
An mtt is finite copying in the parameter if 
there exists a constant $c$ such that
for any $q$, $s$, and $u \in \sem{\tup{q,s}(y_1,\dots,y_k)}$,
the number of occurrences of $y_i$ in $u$ is no more than $c$;
the number $c$ is called a (parameter) {\em copying bound} by $M$.
Note that ``linear-in-parameter'' mtts are a special case of finite copying mtts;
they are not only finite copying with copying bound 1, but also
the finiteness can be known by simply counting the number of syntactic occurrences
of each variable in the rules, while finite copying in general
is a semantic property of mtts.
Also note that finite copying is a decidable property, and the copying bound
can be effectively obtained. (See Lemma~4.10 of~\cite{EM03a}. Although it is proved
only for total deterministic mtts, the same technique also works for IO- and OI-
nondeterministic mtts.)

\begin{theorem} \label{MTTOIfc}
Let $M$ be an mtt that is finite copying in the parameters with copying bound $c$.
Then, translation membership for $\tau\/_{\OI,M}$ can be determined
in time $O(\abs{s} \cdot \abs{t}^{c(2m+2)} \cdot c \cdot \abs{M})$
where $m$ is the maximum rank of $M$'s states.
\end{theorem}
\begin{proof}
Let $t_{\it dag}$ be the minimal dag representing $t$.
Let $V$ be the set of nodes of $t_{\it dag}$.
We define $\labelfn(v)$ to denote the label in $\Sigma$ of the node $v \in V$,
and ${\it child}(v, i)$ to denote the $i$-th child node of $v$.

Let $A = 2^{\bigcup_i Q^{(i)} \times \mathcal{P}_c(V)^i \times V}$
where $\mathcal{P}_c(V) = \{S \subseteq V \mid \abs{S} \leq c \}$
and the function ${\it run}$ be defined as follows:
\[
  {\it run}( \sigma(s_1, \dots, s_k) ) = {\it tr}(
    \sigma, {\it run}(s_1), \dots, {\it run}(s_k)
  ).
\]
The transition function ${\it tr} : (\bigcup_i \Sigma^{(i)} \times A^i) \to A$ is defined as follows
\[
  {\it tr}( \sigma, \vec{a} ) =
   \Big\{
      (q, \vec{\beta}, v') \in \bigcup_i Q^{(i)} \times \mathcal{P}_c(V)^i \times V
      \\\tag*{$\displaystyle %%%% Right-align
    \Big|\ 
        \exists r \in R_{q,\sigma}:
          f_{\vec{\beta}, \vec{a}}(r, v')
   \Big\}
      $} %%%% Right-align
\]
where $f_{\vec{\beta}, \vec{a}} :
T_{\Delta \cup (Q \times X) \cup Y} \times V \to \{{\it true}, {\it false}\}$
defined as follows:
\begin{align*}
f_{\vec{\beta}, \vec{a}}(y_i, v') & = {\it true}  \qquad \,\text{if } v' \in \beta_i \\
f_{\vec{\beta}, \vec{a}}(y_i, v') & = {\it false} \qquad \text{if } v' \not\in \beta_i \\
f_{\vec{\beta}, \vec{a}}(\delta(r_1, \dots, r_n), v') & =
  \bigwedge_{1 \leq i \leq n} \hspace{-6pt} f_{\vec{\beta}, \vec{a}}(r_i, {\it child}(v',i))
  \\
   & \phantom{= \it false} \qquad \ \text{if } \labelfn(v') = \delta  \\
f_{\vec{\beta}, \vec{a}}(\delta(r_1, \dots, r_n), v') & =
  {\it false}
  \qquad \text{if } \labelfn(v') \neq \delta  \\
f_{\vec{\beta}, \vec{a}}(\tup{q', x_j}(r_1, \dots, r_n), v') & = \\
  \exists \vec{\gamma}: (
     (q', \vec{\gamma}, v') \in a_j
     \text{ and }& 
     \text{ for all } i \text{ and } u \in \gamma_i:
     f_{\vec{\beta}, \vec{a}}(r_i, u) 
  ).
\end{align*}
Note that we do not have the $\bot$ element in $V$ this time.
Instead, the empty set $\emptyset$ plays the same role.
The complexity of this algorithm is computed similarly to the case of IO-mtts:
we need to test by $f_{\vec{\beta}, \vec{a}}$
all combinations of $a \in \bigcup_i Q^{(i)} \times \mathcal{P}_c(V)^i \times V$
 (which is of size $O(\abs{Q}\cdot\abs{V}^{cm+1})$ this time)
and $r \in R_{q,\sigma}$,
then $f_{\vec{\beta}, \vec{a}}$ receives $\abs{r} \cdot \abs{V}$ different pairs of arguments,
and finally the computation of $f_{\vec{v}, \vec{a}}(\tup{q',x_j}(\cdots))$
takes $O(\abs{V}^{cm} \cdot c)$ time where $\abs{V}^{cm}$ comes from the part ``$\exists \vec{\gamma}$''
and $c$ comes from the part ``$u \in \gamma_i$''.
The correctness is shown by proving the following claim.

\textit{Claim}\quad 
For every input tree $s'$,
$t' \in \sem{\tup{q,s'}(u_1,\dots,u_n)}^M_\OI$
if and only if
there exist subtrees $t_{1,1},\dots,t_{1,l_1},\dots,t_{n,1},\dots,\allowbreak t_{n,l_n}$ of $t$ such that
 $\{t_{i,1},\dots\hspace{-1.01pt},t_{i,n_i}\} \subseteq \sem{u_i}^M_\OI$ with $l_i\!\leq\!c$
and $(q,\hspace{-1.4pt}(\hspace{-1pt}\{\rho(t_{1,1}),\allowbreak \dots,\rho(t_{1,l_1})\},\dots,
\{\rho(t_{n,1}),\dots,\rho(t_{n,l_n})\}), \allowbreak \rho(t')) \hspace{-1.5pt}\in {\it run}(s')$,
where $\rho$ is defined as in the proof of Theorem~\ref{MTTIO}.

The proof is by induction, too. The finite-copying property ensures that in the
semantics of the mtt, OI-substitution is done only on parameters $y_i$ whose 
number of occurrence
is less than or equal to $c$. It justifies that our algorithm only considers sets of
size $\leq\!c$ as parameter representation.
\end{proof}

%%%%%%%%%%%%%%%%%%%%%%%%%%%%%%%%%%%%%%%%%%%%%%%%%%%%%%%%%%%%%%%%%%%%%%%%%%%%%%%%

On the other hand, the PTIME result for IO-mtts can be generalized to
a more powerful extension of IO-mtts.
One popular way to extend mtts is by {\em regular look-ahead}.
Mtts with regular look-ahead are equipped with one deterministic bottom-up tree automaton
and are allowed to select a rule with respect to the state of the tree automaton,
in addition to the current state and the label of the current node.
Since any $\MTTIO$'s with regular look-ahead
can be simulated by a normal $\MTTIO$ (Theorem 5.19 of~\cite{EV85}),
the translation membership for $\MTTIO$ with regular look-ahead is also in PTIME.
In fact, we can further extend the model to use a more expressive model
of look-ahead, namely, tree automata with equality and disequality constraints~$\cite{BT92}$,
while still preserving the PTIME translation membership.

\begin{definition}
A {\em bottom-up tree automaton with equality and disequality constraints} (TAC)
is a tuple $B = (P,\Sigma,\delta)$, where $P$ is the set of states, $\Sigma$ the input alphabet,
and $\delta$ is a set of transitions of the form
$(\sigma^{(m)}, p_1, \dots, p_m, E, D, p)$
where $E, D \subseteq \{1, \ldots, m\}^2$ are the sets of equality and disequality
constraints, respectively.
A list of trees $t_1, \dots, t_m$ is said to satisfy the constraints if
$\forall (i,j) \in E : t_i = t_j$ and $\forall (i,j) \in D : t_i \neq t_j$.
We define $\tilde{\delta}$ inductively as follows:
\begin{align*}
\tilde{\delta} ( \sigma(t_1, \dots, t_m) )
  = \{p \in P &\mid \\ \exists (\sigma, p_1, \dots, p_m, E, D, &\,p) \in \delta :\\
        p_i \in \tilde{\delta}(t_i) \text{ for all }& i \text{ and }
        t_1, \dots, t_m \text{ satisfy $E$ and \hspace{-1pt}$D$}
    \}.
\end{align*}
A TAC is total and deterministic if for any $\sigma \in \Sigma$, $p_1, \dots, p_m \allowbreak \in P$,
and $t_1, \dots, t_m \in T_\Sigma$, there exists one unique transition
$(\sigma^{(m)}, p_1, \dots, p_m, E, D, p) \in \delta$ such that 
$t_1, \dots, t_m$ satisfies the constraints $E$ and $D$.
For a total deterministic TAC,
we abuse the notation and denote by $\tilde{\delta}(t)$ the unique element of itself.
\end{definition}
Note that, as well as a normal bottom-up tree automaton,
we can run a TAC on a tree in (amortized) linear time, by first computing the minimal
dag representation of the input tree;
due to its minimality, the equality (or disequality) test of two subtrees
can be carried out in constant time, by a single pointer comparison.
Also note that total deterministic TACs are equally expressive as
its nondeterministic version (as shown in Proposition~4.2 of \cite{BT92} by a variant of
usual powerset construction). Hence, we adopt total deterministic TACs as our look-ahead
model for mtts, without sacrificing the expressiveness.
\begin{definition}
An {\em mtt with TAC look-ahead} is a tuple $M = (Q,q_0,\Sigma,\Delta,R,B)$
where $B=(P,\Sigma,\delta)$ is a total and deterministic TAC, and all other components
are defined as for mtts, except that the form of rules are as follows:
$$ \tup{q,\sigma(x_1,\dots,x_k)}(y_1, \dots, y_m) \to r \qquad (p_1,\dots,p_k,E,D). $$
The set of right-hand side of all rules of such form is denoted by
 $R_{q,\sigma,p_1,\dots,p_k,E,D}$.
The size $\abs{M}$ is defined as for normal mtts.
\end{definition}
The semantics of mtts with TAC look-ahead differs from normal mtts only
in the side-condition of state application, which is defined as follows:
\begin{align*}
  \sem{ \tup{q,\sigma(s_1, \ldots, s_k)}(u_1, \ldots, u_m) }^M_\mu
    &=
      \\\tag*{$\displaystyle %%%% Right-align
        \bigcup_{r \in R'} \Big(
          \bigsem{ r[x_1/s_1, \dots, x_k/s_k] }^M_\mu
                \xleftarrow[\mu]{} (\sem{u_1}^M_\mu, \ldots, \sem{ u_m }^M_\mu)
        \Big)
      $} %%%% Right-align
      \\\tag*{$\displaystyle %%%% Right-align
\text{where } R' = R_{q,\sigma,\tilde{\delta}(s_1),\dots,\tilde{\delta}(s_k),E,D}
\text{ such that}
      $} %%%% Right-align
      \\\tag*{$\displaystyle %%%% Right-align
 s_1, \dots, s_k \text{ satisfies } E \text{ and } D.
      $} %%%% Right-align
\end{align*}
In a word, rules in $R_{q,\sigma,p_1,\dots,p_k,E,D}$ are used when
the state $q$ is applied to a node satisfying all the following three conditions:
(1) labeled $\sigma$,
(2) the child subtrees $s_1, \dots, s_k$ of the node satisfy the constraints
$E$ and $D$,
and (3) $\tilde{\delta}(s_i) = p_i$ for all $i$.

Mtts with TAC look-ahead are strictly more expressive than normal mtts.
For example, the translation $\{ (\pi(s,s), e) \mid s \in T_\Sigma \}$
where $\pi$ is a symbol of rank 2 and $e$ is of rank 0,
can be done by a transducer with TAC look-ahead.
But no mtt-composition can realize this translation because
the domain is not regular (by Corollary~5.6 of~\cite{EV85},
the domain of any mtt must be a regular tree language).
Nevertheless, the PTIME translation membership for $\MTTIO$
can be extended to mtts with TAC look-ahead.

\begin{theorem} \label{MTTIOR}
Let $M$
 be an mtt with TAC look-ahead.
Translation membership for $\tau\/_{\IO,M}$ can be determined
in time $O(\abs{s} \cdot \abs{t}^{2m+2} \cdot \abs{M})$ where $m$ is the maximum rank
of $M$'s states.
\end{theorem}
\begin{proof}
The basic idea is again the on-the-fly construction of the inverse-type automaton,
but this time, to deal with the look-ahead, we run parallely the look-ahead automaton.

Let $s_{\it dag}$ be the minimal dag representation of $s$, which can be computed in $O(\abs{s})$ time.
As explained before, the equality (or disequality) test of two subtrees of $s_{\it dag}$
can be carried out in constant time.
Let $V_{s}$ be the set of nodes of $s_{\it dag}$.
Let $V_t$ be the set of nodes of $t_{\it dag}$ and $V = V_t \cup \{\bot\}$.
The functions $\labelfn(v)$,
${\it child}(v, i)$, and $\rho(t)$ are defined as in the proof of Theorem~\ref{MTTIO}.

Let $A = 2^{\bigcup_i Q^{(i)} \times V^i \times V}$ and
${\it run} : T_\Sigma \to V_{s} \times P \times A$ (note the difference
of the return value of ${\it run}$, compared to that in
Theorem~\ref{MTTIO}) be the function defined as follows
\begin{align*}
  {\it run}( s' ) &= {\it tr}(s',
    \sigma, {\it run}(s_1), \dots, {\it run}(s_k)
  )
        \\\tag*{$\displaystyle %%%% Right-align
  \text{ with } s' = \sigma(s_1, \dots, s_k)
        $} %%%% Right-align
\end{align*}
where the function {\it tr} is:
\begin{align*}
   {\it tr}&( s', \sigma, (s_1,p_1,a_1),\dots,(s_k,p_k,a_k) ) =
 \\
   &\Big(
      s',
      \tilde{\delta}(s'), 
\\
      &\ \big\{
        (q, \vec{v}, v') \in \bigcup_i Q^{(i)} \times V^i \times V
      \,\big|\,
        \exists r \in R_{q,\sigma,p_1,\dots,p_k,E,D}:
        \\[-0.5em]\tag*{$\displaystyle %%%% Right-align
            (s_1,\dots,s_k) \text{ satisfies } E, D
            \text{ and } f_{\vec{v}, \vec{a}}(r, v')
     \big\}
   \Big).
        $} %%%% Right-align
\end{align*}
The definition of $f_{\vec{v}, \vec{a}}$ remains the same as in Theorem~\ref{MTTIO}.

The look-ahead state $\tilde{\delta}(s')$ can be computed from
$\sigma$, $p_1, \dots, \allowbreak p_k$, and $s_1, \dots, s_k$ in constant time.
By the same argument as the case of normal mtts,
 we obtain the $O(\abs{s} \cdot \abs{t}^{2m+2} \cdot \abs{M})$ time complexity.
The correctness of the construction is proved also in the same way as for normal mtts.
That is, we can prove the following claim
by nested induction on structure of $s'$, and then on the structure of
right-hand sides of the rules.

\textit{Claim}\quad 
For every input tree $s'$, we have the following equation for all
 $q \in Q$, 
 $r_i \in T_{\Delta \cup (Q \times T_\Sigma) \cup Y}$, and an environment $\Gamma$:
$
\rho\Big( \sem{\tup{q,s'}(r_1,\dots,r_n)}^M_\IO \Big)
=
 \Big\{
   v' \ \Big|\  (q,(v_1,\ldots,v_n),v') \in {\it run}(s'), v_i \in \rho(\sem{r_i}^M_\IO)
   \text{ for all }i
 \Big\}
$

Again, applying the claim to $\rho(\sem{\tup{q_0,s}}^M_\IO)$,
we know that the translation membership is equivalent to
 $(q_0,(),v_\epsilon) \in {\it run}(s)$ where $v_\epsilon$ is the root node of $t_{\it dag}$.
Hence, the translation membership can be determined by computing the set ${\it run}(s)$.
\end{proof}

%%%%%%%%%%%%%%%%%%%%%%%%%%%%%%%%%%%%%%%%%%%%%%%%%%%%%%%%%%%%%%%%%%%%%%%%%%%%%%

Another extension of mtts that admits a polynomial time translation
membership is {\em multi-return mtts (mr-mtts\/)}~\cite{IH07,IHM08}.
In an mr-mtt, states may return multiple trees (with the initial
state returning exactly one tree).
Mr-mtts are strictly more expressive than normal mtts,
and furthermore, have better closure properties under composition with
top-down tree transducers~\cite{IHM08}.

\begin{definition}
A {\em multi-return macro tree transducer (mtt)} $M$
is a tuple $(Q,\Sigma,\Delta,q_0,R,D)$,
where $Q$, $\Sigma$, $\Delta$, and $q_0$ are defined as for mtts,
$D : Q \to \nat$  is the {\em dimension} such that $D(q_0) = 1$, and $R$
is the finite set of {\em rules} of the form
$$ \tup{q, \sigma(x_1,\dots,x_k)} (y_1, \allowbreak \ldots, y_m)
   \to r $$
where $q \in Q^{(m)}$, $\sigma \in \Sigma^{(k)}$,
 and  $r \in \rhs^{D(q)}_{X_k}$\! where for $e \geq 1$ and a set $Q$, the set $\rhs^e_W$ is defined as:
   \begin{align*}
    r &::= l_1\ldots l_n \ (u_1, \dots, u_e) \hspace{110pt} (n \geq 0)
  \\
    l &::= \text{let } (z_{j+1}, \dots, z_{j+D(q')}) \text{ = }
       \tup{q', x_i} (u_1, \dots, u_n) \text{ in}
      \\\tag*{$\displaystyle %%%% Right-align
        (j \in \nat, q' \in Q^{(n)}, x_i \in W)
      $} %%%% Right-align
   \end{align*}
  with $u_1, u_2, \ldots \in T_{\Delta \cup Y_m \cup Z}$.
  We usually omit parentheses around tuples of size one, i.e., write like
  $\mbox{let\,} z_j \mbox{\,=\,} \cdots \mbox{in\,} u_1$.
  We require any rule to be well-formed, that is, the leftmost occurrence of any variable $z_i$
  must appear at a ``binding'' position (between `let' and `='), and the next occurrence (if any)
  must appear after the `in' corresponding to the binding occurrence.
  The set of right-hand sides of such rules is denoted by $R_{q,\sigma}$.
  The size $\abs{M}$ of the mr-mtt is defined to be the sum of the size of right-hand sides,
  i.e., the number of $\delta$, $Y$, $Z$, and $Q \times X$ nodes.
\end{definition}

The IO-semantics of mr-mtts is inductively defined as follows.
For $u \in T_{\Delta \cup Y_m \cup Z}$, $\sem{u}^M_\IO \subseteq T_{\Delta \cup Y \cup Z}$ is
\begin{align*}
  \sem{ \delta(u_1, \dots, u_k) }^M_\IO
    &=
       \{ \delta(t_1, \ldots, t_e) \mid t_i \in \sem{u_i}^M_\IO \text{ for all } i \}
\\
  \sem{ y_i }^M_\IO
    &=
       \{ y_i \}
\\
  \sem{ z_i }^M_\IO
    &=
       \{ z_i \}
\end{align*}
and for $\kappa \in \rhs^e_{T_\Sigma}$, $\sem{\kappa}^M_\IO \subseteq T_{\Delta \cup Y \cup Z}^e$ is
\[
\begin{array}{l}
  \sem{ (u_1, \dots, u_e) }^M_\IO
    =
       \{ (t_1, \ldots, t_e) \mid t_i \in \sem{u_i}^M_\IO \text{ for all } i \}
\\[3pt]
  \sem{ \text{let }(z_1,\dots,z_d) = \tup{q,\sigma(s_1,\dots,s_m)}(u_1, \dots, u_k) \text{ in } \kappa'
  }^M_\IO
\\
\displaystyle
  = \Big\{
       \xi[z_1/t_1, \ldots, z_d/t_d]
    \mid
\\\hspace{16pt}
       \xi \in \sem{\kappa'}^M_\IO,
       r \in R_{q,\sigma},
       (t_1, \ldots, t_d) \in
\\[-10pt]
      \tag*{$\displaystyle %%%% Right-align
   \Big(
    \sem{ r[x_1/s_1, \dots, x_m/s_m] }^M_\IO
          \xleftarrow[\rm IO]{} (\sem{ u_1 }^M_\IO, \ldots, \sem{ u_k }^M_\IO)
   \Big)
  \Big\}.
      $} %%%% Right-align
\end{array}
\]
The translation $\tau\/_{\IO,M} \subseteq T_\Sigma \times T_\Delta$ realized by $M$
is the set $\{(s,t) \mid t \in \sem{\text{let } z = \tup{q_0,s} \text{ in } z}\}^M_\IO$.

Here is an example of an mr-mtt, which is used in~\cite{IH07} as a counterexample
that cannot be realized in normal mtts:
\begin{align*}
  \tup{ q_0, {\tt s}(x) } () &\to\text{let}\ (z_1, z_2) =
     \tup{ q_1, x } ({\tt A}({\tt E}))\ \text{in}\ {\tt r}({\tt a}(z_1), z_2) \\
  \tup{ q_0, {\tt s}(x) } () &\to\text{let}\ (z_1, z_2) =
     \tup{ q_1, x } ({\tt B}({\tt E}))\ \text{in}\ {\tt r}({\tt b}(z_1), z_2) \\
  \tup{ q_0, {\tt z} } () &\to{\tt r}({\tt e}, {\tt E}) \\
  \tup{ q_1, {\tt s}(x) } (y_2) &\to\text{let}\ (z_1, z_2) =
     \tup{ q_1, x } ( {\tt A}(y_2))\ \text{in}\ ({\tt a}(z_1), z_2) \\
  \tup{ q_1, {\tt s}(x) } (y_2) &\to\text{let}\ (z_1, z_2) =
     \tup{ q_1, x } ( {\tt B}(y_2))\ \text{in}\ ({\tt b}(z_1), z_2) \\
  \tup{ q_1, {\tt z} } (y_2) &\to({\tt e}, y_2)
\end{align*}
This nondeterministic translation takes as input
monadic trees of the form ${\tt s}({\tt s}(\cdots {\tt s}({\tt z})\cdots))$
and produces output trees of the form  ${\tt r}(t_1, t_2)$ where
    $t_1$ is a monadic tree over ${\tt a}$'s and ${\tt b}$'s (and a leaf ${\tt e}$),
and $t_2$ is a monadic tree over ${\tt A}$'s and ${\tt B}$'s such that
$t_2$ is the reverse of $t_1$, and both have the same size as the input.
For instance,
  ${\tt r}({\tt a}({\tt a}({\tt b}({\tt e}))),
            {\tt B}({\tt A}({\tt A}({\tt E}))))$
is a possible output tree for the input
           ${\tt s}({\tt s}({\tt s}({\tt z})))$.
Consider the return value of the state call
$\sem{\tup{q_1,{\tt s}({\tt z})}({\tt E})}$:
it is the set $\{
  ({\tt a}({\tt E}), {\tt A}({\tt E})),
  ({\tt b}({\tt E}), {\tt B}({\tt E}))
\}$ of {\em pairs of} trees.
In a word, the state $q_1$ returns only {\em mutually reverse} pairs of monadic trees.
This is impossible in normal mtts,
 in which we must carry out two state calls in order to obtain two output trees;
 two nondeterministic state calls are evaluated independently, and cannot avoid
 generating unrelated pairs of trees.

Despite their expressive power over normal mtts, mr-mtts still have
a similar complexity for inverse type inference. Therefore 
the translation membership remains in PTIME.

\begin{theorem} \label{MRMTT}
Let $M$ be an mr-mtt.
Translation membership for $\tau\/_{\IO,M}$ can be determined
in time $O(\abs{s} \cdot \abs{t}^{2m+2d} \cdot \abs{M})$
where $m$ is the maximum rank of the states and
$d$ is the maximum dimension.
\end{theorem}
\begin{proof}
For mr-mtts, we take the set $A$ of inverse-type automaton
as $A = 2^{\bigcup_{i,j} Q^{(i,j)} \times V^i \times V^j}$
where $Q^{(i,j)}$ is the set of states $q$ of $\rankfn(q)=i$ and $D(q)=j$.
The intuition of the set of states $A$ is similar to the case of normal mtts.
That is,
``$(q,\vec{v},\vec{w}) \in {\it run}(s')$'' means that ``if $q$ is applied to
the input subtree $s'$ with output subtrees rooted at $\vec{v}$ as parameters,
then it may return a tuple of output subtrees $\vec{w}$''.
The construction is quite similar to that of the proof of Theorem~\ref{MTTIO}.
\end{proof}

As a final remark we would like to mention the complexity of
translation membership for deterministic mtts;
it can be determined in linear time.
Since domains of compositions of mtts are regular,
we can factor out the partiality and have the following decomposition:
for $\mu \in \{\IO, \OI\}$,
 $\DMTT_\mu^n \subseteq {\rm FTA} \comp \DtMTT^n$ where ${\rm FTA}$ is the class of
partial identities whose domain is regular (analogous to Theorem 6.18 of~\cite{EV85}).
Therefore, to compute the translation membership for a composition of deterministic mtts,
we first check in $O(\abs{s})$ time whether the given input $s$ is contained in the domain of
the translation, and then check the translation membership for composition of deterministic 
{\em and total\/} mtts.
Here, by Theorem~15 of~\cite{Maneth02},
for a translation $\tau \in \DtMTT^n$
we can compute the unique output tree $t' \in \tau(s)$ from the input $s$ in time $O(\abs{s}+\abs{t'})$,
and during the computation, the size of every intermediate tree is less than or equal to $2^n \cdot \abs{t'}$.
Hence, for testing $(s,t) \in \tau$, we simply compute $\tau(s)$;
if the size of any intermediate tree exceeds $2^n \cdot \abs{t}$ then $(s,t)$ cannot be an element of
$\tau$, and otherwise, we compare the computed tree $\tau(s)$ with $t$.
The time complexity of the above procedure is $O(\abs{s} + 2^n \cdot \abs{t})$.

\begin{theorem} \label{DMTT}
Let $\mu \in \{\IO,\OI\}$ and $n \geq 1$. Translation membership for 
$\DMTT^n_\mu$ is in $O(\abs{s}+2^n \abs{t})$.
\end{theorem}

%%%%%%%%%%%%%%%%%%%%%%%%%%%%%%%%%%%%%%%%%%%%%%%%%%%%%%%%%%%%%%%%%%%%%%%%%%%%%%%%

\section{Future Work}
\label{sec:future}

The complexity of the translation membership problem remains open for
several interesting subclasses and extensions of mtts.
One example is the mtt with holes~\cite{MN08} in IO mode.
Note that, similar to Theorem~4.6 of~\cite{MN08},
hole-mtts in IO mode are equal to $\MTTIO \comp \YIELD$,
which is included in $\MTTIO \comp \LDtMTT$.
An algorithm based on inverse type inference does not work, because the
parameter part of the states of the inverse-type automaton is a set of
functions $[V \to V]$, which is exponential in size with respect
to the output tree $\abs{t}$.
On the other hand, it is not clear either whether it is NP-hard.
Note that mtts with holes in OI mode can simulate all OI-mtts,
and therefore their translation membership is NP-complete.

Another interesting class is that of {\em 1-parameter} mtts in OI mode.
Our encoding of 3-SAT used three parameters.
In fact, the number of parameters can be reduced to two by embedding
the encodings of boolean variables in the input tree $s$.
Can we encode 3-SAT into a 1-parameter mtt? Or, do
1-parameter mtts actually have PTIME translation membership?
(Again, the inverse-type automaton technique used in this
paper for IO-mtts does not seem to work in this case,
because the automaton gets too large.)

%%%%%%%%%%%%%%%%%%%%%%%%%%%%%%%%%%%%%%%%%%%%%%%%%%%%%%%%%%%%%%%%%%%%%%%%%%%%

\textbf{Acknowledgments}
This work was partly supported by
Japan Society for the Promotion of Science.

\bibliographystyle{abbrv}
\bibliography{my}

\begin{thebibliography}{10}

\bibitem{BT92}
B.~Bogaert and S.~Tison.
\newblock Equality and disequality constraints on direct subterms in tree
  automata.
\newblock In {\em Symposium on Theoretical Aspects of Computer Science
  (STACS)}, 1992.

\bibitem{Courcelle94}
B.~Courcelle.
\newblock Monadic second-order definable graph transductions: A survey.
\newblock {\em Theoretical Computer Science}, 126:53--75, 1994.

\bibitem{DST80}
P.~J. Downey, R.~Sethi, and R.~E. Tarjan.
\newblock Variations on the common subexpression problem.
\newblock {\em Journal of the ACM}, 27:758--771, 1980.

\bibitem{EM03}
J.~Engelfriet and S.~Maneth.
\newblock A comparison of pebble tree transducers with macro tree transducers.
\newblock {\em Acta Informatica}, 39:613--698, 2003.

\bibitem{EM03a}
J.~Engelfriet and S.~Maneth.
\newblock Macro tree translations of linear size increase are mso definable.
\newblock {\em SIAM Journal on Computing}, 32:950--1006, 2003.

\bibitem{EV85}
J.~Engelfriet and H.~Vogler.
\newblock Macro tree transducers.
\newblock {\em Journal of Computer and System Sciences}, 31:71--146, 1985.

\bibitem{Fischer68}
M.~J. Fischer.
\newblock {\em Grammars with Macro-Like Productions}.
\newblock PhD thesis, Harvard University, Cambridge, 1968.

\bibitem{GareyJohnson}
M.~R. Garey and D.~S. Johnson.
\newblock {\em Computers and Intractability: A Guide to the Theory of
  {NP-Completeness}}.
\newblock Freeman, 1979.

\bibitem{IH07}
K.~Inaba and H.~Hosoya.
\newblock {XML} transformation language based on monadic second order logic.
\newblock In {\em Programming Language Technologies for XML (PLAN-X)}, pages
  49--60, 2007.

\bibitem{IHM08}
K.~Inaba, H.~Hosoya, and S.~Maneth.
\newblock Multi-return macro tree transducers.
\newblock In {\em Conference on Implementation and Application of Automata
  (CIAA)}, 2008.

\bibitem{IM08}
K.~Inaba and S.~Maneth.
\newblock The complexity of tree transducer output languages.
\newblock In {\em Foundations of Software Technology and Theoretical Computer
  Science (FSTTCS)}, 2008 (Available at
  \url{http://arbre.is.s.u-tokyo.ac.jp/~kinaba/fst.pdf}).

\bibitem{Maneth02}
S.~Maneth.
\newblock The complexity of compositions of deterministic tree transducers.
\newblock In {\em Foundations of Software Technology and Theoretical Computer
  Science (FSTTCS)}, 2002.

\bibitem{MBPS05}
S.~Maneth, A.~Berlea, T.~Perst, and H.~Seidl.
\newblock {XML} type checking with macro tree transducers.
\newblock In {\em Principles of Database Systems (PODS)}, 2005.

\bibitem{MN08}
S.~Maneth and K.~Nakano.
\newblock {XML} type checking for macro tree transducers with holes.
\newblock In {\em Programming Language Technologies for XML (PLAN-X)}, 2008.

\bibitem{MPS07}
S.~Maneth, T.~Perst, and H.~Seidl.
\newblock Exact {XML} type checking in polynomial time.
\newblock In {\em International Conference on Database Theory (ICDT)}, 2007.

\bibitem{MSV03}
T.~Milo, D.~Suciu, and V.~Vianu.
\newblock Typechecking for {XML} transformers.
\newblock {\em Journal of Computer and System Sciences}, 66:66--97, 2003.

\bibitem{PS04}
T.~Perst and H.~Seidl.
\newblock Macro forest transducers.
\newblock {\em Information Processing Letters}, 89:141--149, 2004.

\bibitem{Rounds73}
W.~C. Rounds.
\newblock Complexity of recognition in intermediate-level languages.
\newblock In {\em Foundations of Computer Science (FOCS)}, 1973.

\end{thebibliography}
\balancecolumns
\end{document}